\documentclass[12pt,preprint]{aa}
\usepackage{graphicx}
\usepackage{txfonts}
\usepackage{CJK}
\usepackage{xcolor}
\usepackage{multirow}
\begin{document}
\begin{CJK*}{UTF8}{gbsn}
\title{Quiescent photometric modulations of two low-inclination cataclysmic variables \object{KZ\,Gem} and \object{TW\,Vir}}
\titlerunning{Photometric modulations of \object{KZ\,Gem} and \object{TW\,Vir}}
\author{Zhibin Dai\inst{1,2,3}, Paula Szkody\inst{4}, Ali Taani\inst{5}, Peter M. Garnavich\inst{6} \and Mark Kennedy\inst{6,7}}
\authorrunning{Z. B. Dai, P. Szkody, A. Taani, P. M. Garnavich, M. Kennedy}
\institute{
Yunnan Observatories, Chinese Academy of Sciences, 396 Yangfangwang, Guandu District, Kunming, 650216, P. R. China.\\
\email{zhibin\_dai@ynao.ac.cn}
\and
Key Laboratory for the Structure and Evolution of Celestial Objects, Chinese Academy of Sciences, 396 Yangfangwang, Guandu District, Kunming, 650216, P. R. China.\\
\and
Center for Astronomical Mega-Science, Chinese Academy of Sciences, 20A Datun Road, Chaoyang District, Beijing, 100012, P. R. China.\\
\and
University of Washington, Seattle, WA, 98195, USA.\\
\and
Applied Sciences Department, Aqaba University College, Al-Balqa' Applied University, P.O. Box 1199, 77110 Aqaba, Jordan.\\
\and
University of Notre Dame, Notre Dame, IN, 46556, USA.\\
\and
Department of Physics, University College Cork, Cork, Ireland.
}
\date{Received / Accepted}

\abstract{}
{The quiescent periodic photometric modulations of two low-inclination cataclysmic variables observed in Kepler K2 Campaigns 0 and 1, \object{KZ\,Gem} and \object{TW\,Vir}, are investigated.}
{A phase-correcting method was successfully used to detect the orbital modulations of \object{KZ\,Gem} and \object{TW\,Vir} and improve their orbital periods. The light curve morphologies of both CVs were further analyzed by defining flux ratios and creating colormaps.} 
{\object{KZ\,Gem} shows ellipsoidal modulations with an orbital period of 0.22242(1)\,day, twice the period listed in the updated RK catalogue (Edition 7.24). With this newly determined period, \object{KZ\,Gem} is no longer a CV in the period gap, but a long-period CV. A part of the quiescent light curve of \object{TW\,Vir} that had the highest stability was used to deduce its improved orbital period of 0.182682(3)\,day. The flat patterns shown in the colormaps of the flux ratios for \object{KZ\,Gem} demonstrate the stability of their orbital modulations, while \object{TW\,Vir} show variable orbital modulations during the K2 datasets. In \object{TW\,Vir}, the single versus double-peaked nature of the quiescent orbital variations before and after superoutburst may be related to the effect of the superoutburst on the accretion disk.}
{}


\maketitle

\section{Introduction}
 
Cataclysmic variables (hereafter CVs) are close binary systems in which a white dwarf (WD) primary accretes matter from a Roche-lobe filling late-type star via the inner Lagrangian point (CVs are reviewed in \cite{war95}). Unless the magnetic field of the white dwarf is in the MG regime, the matter accumulates in an accretion disk surrounding the white dwarf and a hot spot is created where the mass transfer stream intersects the accretion disk. The low-inclination CVs show no eclipses and lack easily detected photometric orbital modulations caused by a changing view of the hot spot during their orbit. To detect small ($<$\,1\,mag) optical modulation features requires high signal-to-noise and a long time base \citep[e.g.][]{szk92,tay99,pat03}. Hence, it is difficult to find suitable data on many low-inclination CVs at quiescence. Considering that flickering is one of the most striking photometric characteristics common to all CVs and has an amplitude of a few hundredths of a magnitude up to more than an entire magnitude \citep{bru91,bru92}, the low-amplitude orbital modulations in faint low-inclination CVs may be completely overwhelmed. For example, data on \object{QZ\,Vir} \citep{dai16} do not show a distinct orbital modulation due to possible large-amplitude flickering. The large scatter shown in the phased light curves of \object{RZ\,Leo} and \object{FO\,Aqr} derived by \cite{dai16} and \cite{ken16}, respectively, indicates that effects from the white dwarf spin can also complicate the extraction of an orbital modulation.

Comparison of light curve data obtained at different times shows that flickering and orbital modulations are variable and the appearance in the measured light curve obviously depends on the exposure times and binning of the data \citep{szk16,dai16}. The unprecedented light curves from the Kepler K2 mission \citep{how14}, with nearly continuous photometric coverage for 1-3 months at different pointings (Campaigns) along the ecliptic provide an excellent database to study low-amplitude photometric variations in faint low-inclination CVs. K2 Campaign 0 (K2-C0) was an engineering test in the early stage of the K2 program and the telescope did not have pointing stability during the 35-day campaign. K2 Campaign 1 (K2-C1) was the first campaign with fine pointing and it covers 80 days.

In this paper, we analyze two CV light curves in detail \object{KZ\,Gem} was observed in long cadence (LC; 30\,min sampling) during K2-C0 and and \object{TW\,Vir} in short cadence (SC; 1\,min sampling) during K2-C1. Both CVs listed in Table 1 are considered low-inclination systems due to the lack of detected eclipses. We apply a phase-correcting method to the quiescent light curves of \object{KZ\,Gem} and \object{TW\,Vir} to refine their orbital periods and study their orbital modulations.

\begin{table}
\caption{The two low-inclination CVs.}
\begin{center}
\begin{tabular}{ccc}
\hline\hline
Name$^{a}$ & \object{KZ\,Gem} & \object{TW\,Vir}\\
\hline
EPIC & 202061320 & 201185922\\
Campaign (LC/SC)$^{c}$ & C0 (LC) &  C1 (SC)\\
Duration (day) & 33.1 & 80.1\\
P$_{orb}(hr)^{b}$ & 2.67 & 4.38\\
Magnitude & V14.7 - 16 & V12.0 - 17.2\\
Type & DN & DN\\
\hline\hline
\end{tabular}
\end{center}
\tablefoot{Note, $^{a}$ the abbreviated names are used for the objects without standard variable star designations. $^{b}$ listed in the CV catalogue (RKcat Edition 7.24). $^{c}$ LC and SC mean the observation in long (30\,min sampling) and short (1\,min sampling) cadence, respectively.}
\end{table}

\section{K2 Light Curve Extractions}

The K2 data are stored in the original Target Pixel Files (TPFs) provided by the Mikulski Archive for Space Telescopes (MAST). \cite{dai16} presented the light curve extractions for \object{TW\,Vir}, but there was only a featureless light curve for the faint CV \object{KZ\,Gem}, extracted by using the program PyKE developed by the Guest Observer Office \citep{sti12} and the pipeline presented by \cite{dai16}. Since the "self-flat-fielding" (SFF) method proposed by \cite{van14a} and \cite{van14b} was claimed to improve the K2 light curves by removing the effect of the trends, the SFF corrected light curves of \object{KZ\,Gem} was used for our further analysis. Note that the SFF corrected data are described by the normalized flux. Although the K2 light curve shown in Figure 1 does not show any visible periodic modulation, the results from Lomb-Scargle periodogram (LSP; \cite{lom76} and \cite{sca82}) and phase dispersion minimization (PDM; \cite{ste78}) found a modulation period from its K2 data.

\begin{figure}
\centering
\includegraphics[width=9.0cm]{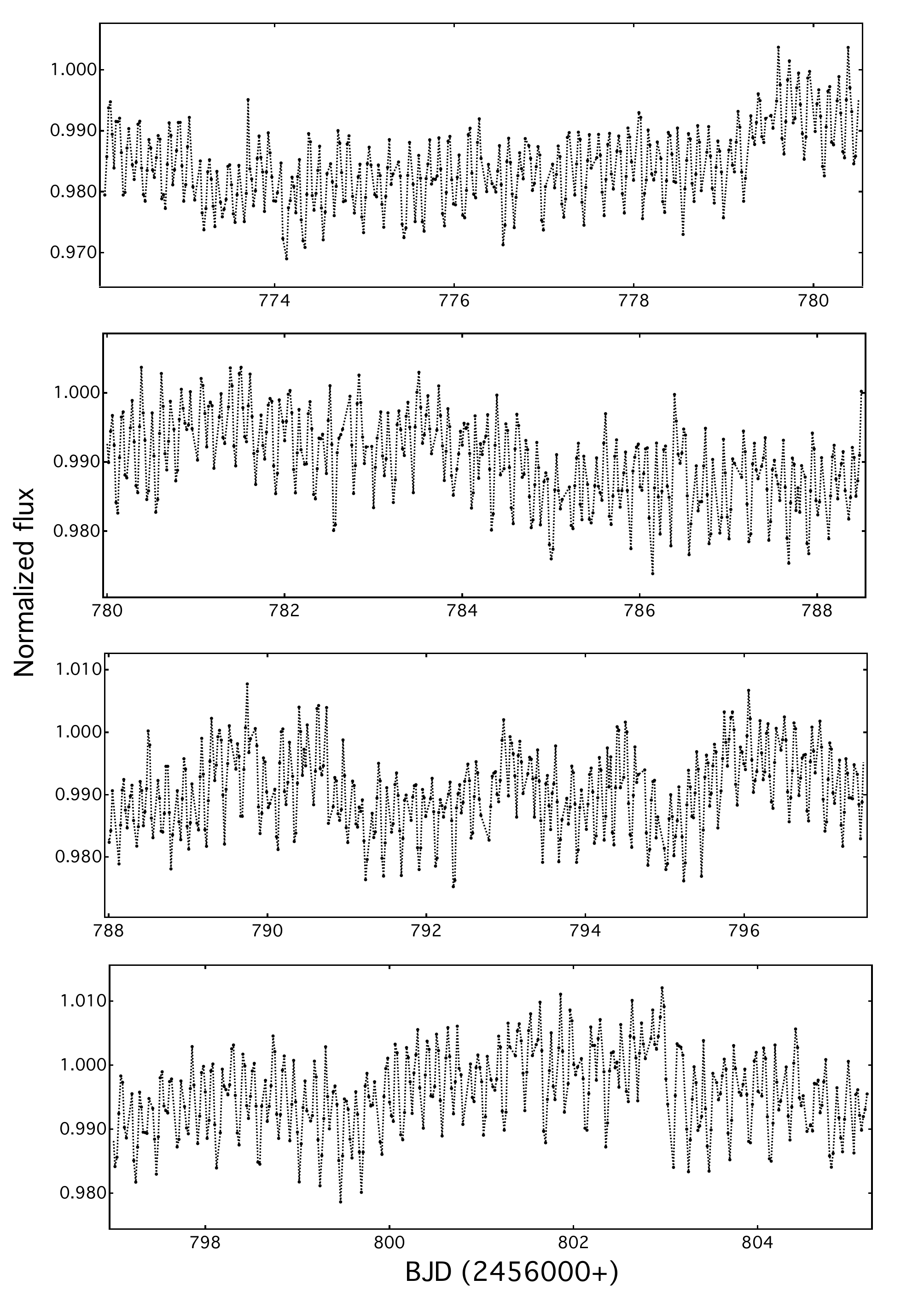}
\caption{The SFF corrected light curve of \object{KZ\,Gem} observed in K2-C0.} \label{Figure 1}
\end{figure}

\section{Analysis Methods}

\subsection{Phase-correcting Method}

The methods of LSP and PDM are generally used to search for periods in time series data, but they are not ideal for finding variable or unstable modulations. Although a 2-D power spectrum analysis can help with this problem, it may only show a rough period with low precision. Thus, based on a preliminary period P$_{0}$ derived from the above period finding methods or an existing radial velocity result, we attempted to apply a phase-correcting method to improve the precision of the observed period. Since P$_{0}$ is a key parameter for the success of our phase-correcting method, it is necessary to make a precise estimation as soon as possible by using the LSP and PDM, which are powerful to find periods in data sets with highly non-regular timing.

The K2 continuous time series data were separated into several sections for phasing the light curve on the starting period P$_{0}$, with each section maintaining the same data length, N$_{s}$, which is measured in units of the cycles. A standard phase (e.g., the light minimum at phase 0 or maximum at phase 0.5) was fixed to be a reference phase. Based on this reference phase, the small and imperceptible discrepancy from the starting period P$_{0}$ (the average phase uncertainty is on the order of tens of seconds for both CVs discussed in this paper) can then be accumulated and amplified to be a measurable parameter accompanying the increased cycles. In principle, the shifting phases of all sections show a linear curve with a constant shift rate S$_{phase}$ derived by a decent linear fit. This discrepancy can be deduced as $\frac{S_{phase}}{N_{s}}\,P_{0}$. The resulting modulation period P$_{b}$ can be described by the following formula, i.e.,
\begin{equation}
P_{b}\,=\,(1-\frac{S_{phase}}{N_{s}})\,P_{0}.
\end{equation}
The data length N$_{s}$ is a key parameter for this phase-correcting method. An appropriate N$_{s}$ should not only satisfy the minimal length for folding the light curve, but also guarantee enough number of sections for deducing the shift rate S$_{phase}$, which determines the precision of the final measured modulation period P$_{b}$. After several tries, we determined that the best N$_{s}$ should be 10 and 5 for the LC and SC data, respectively. The orbital periods of the two CVs improved by using this method are discussed in detail in Section 4.

This method is suitable for analyzing data where the periodic variation is close to a simple harmonic model. Particularly, it is good at accurately finding modulations which undergo amplitude changes during a long data string. By separating a complete time series into many sections, it is easy to figure out when the modulations begin to become unstable, indicated by the first chaotic section. Thus, the part of the light curve with the highest stability can be picked out to derive the modulation period by using this method. And the modulation can be visualized by phasing the stable part of the light curve. \object{TW\,Vir} as a good example is presented at length in Section 4.2.1. Note that this method depends on an implicit assumption that the phase of the variability is not intrinsically variable during the time span of the data.  

\subsection{Flux Ratios}

For a double-hump light curve, the hump at the higher flux level is defined as the primary peak, and the lower flux hump is the secondary peak. Similarly, the light minimum at the lower flux level is the primary dip, and the other minimum is the secondary dip. Taking advantage of the continuous high-precision K2 data, these peaks and dips can be used to describe the light-curve morphologies. Although the K2 data are not absolutely calibrated, the relative flux changes with time can be determined. We use four flux ratios that are defined by the following formulas,
\begin{equation}
\left \{
\begin{array} {l}
R_{peak}\,=\,f_{sp}\,/\,f_{pp}\\
R_{dip}\,=\,f_{sd}\,/\,f_{pd}\\
A_{pri}\,=\,f_{pd}\,/\,f_{pp}\\
A_{sec}\,=\,f_{sd}\,/\,f_{sp}\\
\end{array}
\right.
\end{equation}
where f$_{sp}$, f$_{sd}$, f$_{pp}$ and f$_{pd}$ are the flux at the secondary peak, secondary dip, primary peak and primary dip, respectively. An accurate modulation period and a parabolic least-square fitting method are used to extract the flux at the peaks and dips. The latter can largely eliminate the effects from discrepant data points and possible random drifts with time. R$_{peak}$ and R$_{dip}$ indicate the flux ratios of two peaks and dips, respectively. A$_{pri}$ and A$_{sec}$ denote the primary and secondary amplitudes of the modulations, respectively. Since both a strengthening of a peak flux or a weakening of a dip flux can give rise to a variation in the hump amplitude, the two parameters A$_{pri}$ and A$_{sec}$ are not enough to completely describe the variations in the light curve. By combining these two ratios with the other two flux ratios R$_{peak}$ and R$_{dip}$, the variations in peaks and dips can be distinguished. This method is mainly considers the effects on the flux from cycle-to-cycle variations, flickering, and observational errors.

For visualization, each flux ratio was plotted into a 25$\times$25\,pix colormap (i.e., a diagram of flux ratio versus BJD). The K2 data were separated into 25 sections and all flux ratios were calculated for each section. The maximum and minimum flux ratio for all 25 sections were then set to be the upper and lower border of the diagrams. Between the two borders, the flux ratio was separated into 25 uniform sections. The number of calculated flux ratios falling into each grid was then counted. The resulting colormaps are shown in Figure 2. Note that large, detectable amplitudes in the primary or secondary hump lead to A$_{pri}$ and A$_{sec} <$ 1, respectively. Also, unchanged flux ratios over the course of the observations indicate a stable modulation. Table 2 lists the averages of the four flux ratios for all 5 targets and the corresponding coefficients of the fits.

\begin{table*}
\caption{The four flux ratios.}
\begin{center}
\begin{tabular}{ccccccc}
\hline\hline
\multicolumn{2}{c}{$^{a}$Flux ratio} & \object{J0632+2536} & \object{KZ\,Gem} & \object{RZ\,Leo} & \object{WD\,1144+011} & \object{TW\,Vir}\\
\hline
\multirow{2}{*}{R$_{peak}$} & avg & 0.95(4) & 1.000(4) & 0.84(4) & -- & --\\
& coef & -1.5(2)$\times$10$^{-3}$ & 7(2)$\times$10$^{-5}$ & 7(4)$\times$10$^{-5}$ & -- & --\\\\
\multirow{2}{*}{R$_{dip}$} & avg & 1.14(4) & 0.999(4) & 1.06(5) & -- & --\\
& coef & $^{b}$2.0(3)$\times$10$^{-4}$ & -4(2)$\times$10$^{-5}$ & -7(5)$\times$10$^{-5}$ & -- & --\\\\
\multirow{2}{*}{A$_{pri}$} & avg & 0.76(3) & 0.996(4) & 0.68(3) & 0.954(6) & 0.80(3)\\
& coef & -9(2)$\times$10$^{-4}$ & 4(3)$\times$10$^{-5}$ & -1.0(3)$\times$10$^{-4}$ & 4(1)$\times$10$^{-5}$ & $^{b}$3.2(7)$\times$10$^{-4}$\\\\
\multirow{2}{*}{A$_{sec}$} & avg & 0.91(2) & 0.994(4) & 0.86(5) & -- & --\\
& coef & 4(1)$\times$10$^{-4}$ & -9(2)$\times$10$^{-5}$ & -2.2(5)$\times$10$^{-4}$ & -- & --\\
\hline\hline
\end{tabular}
\end{center}
\footnotesize{Note, $^{a}$ the average and the best-fitting coefficient of flux ratio are denoted by the symbols of avg and coef, respectively. $^{b}$ it is not the gradient, but the quadratic coefficient.}
\end{table*}

\begin{figure*}
\centering
\includegraphics[width=15.0cm]{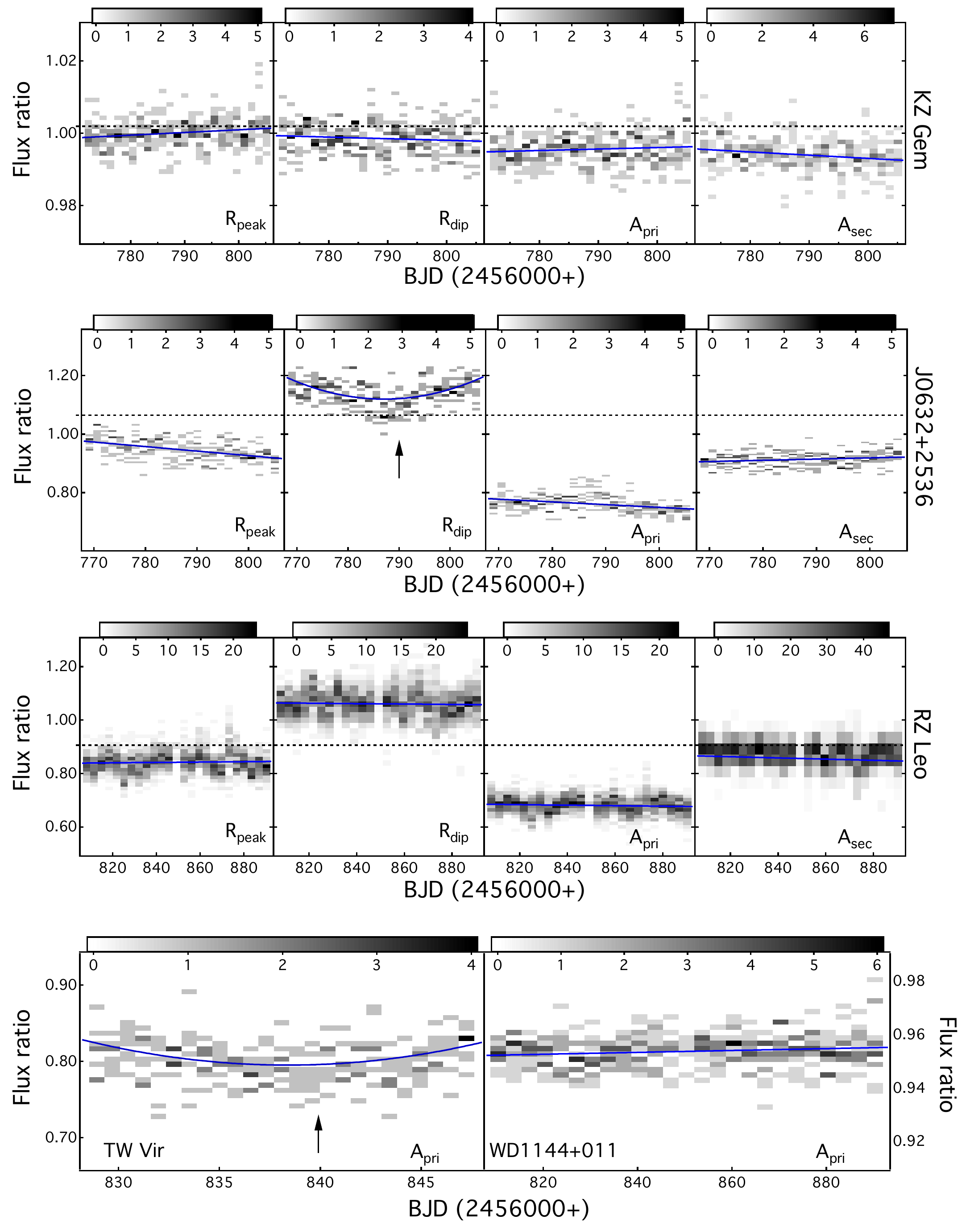}
\caption{The 25$\times$25\,pix colormaps of four flux ratios defined by Equation 2 for the same object are aligned from top to bottom. The two plots of A$_{pri}$ for \object{TW\,Vir} and \object{WD\,1144+011} lie alongside at the bottom, since A$_{pri}$ is the only available parameter for both CVs. The gray scale denotes the number of the flux ratios falling into each grid. The blue solid lines refer to the best-fitting curves, and the average and the corresponding coefficients of flux ratios are listed in Table 2. The arrow shown in R$_{dip}$ plot of \object{J0632+2536} indicates the transient event of \object{J0632+2536} around BJD\,2456790. The time pointed by the arrow in A$_{pri}$ plot of \object{TW\,Vir} is the same as that in panel a) of Figure 8.} \label{Figure 2}
\end{figure*}

\begin{figure}
\centering
\includegraphics[width=9.0cm]{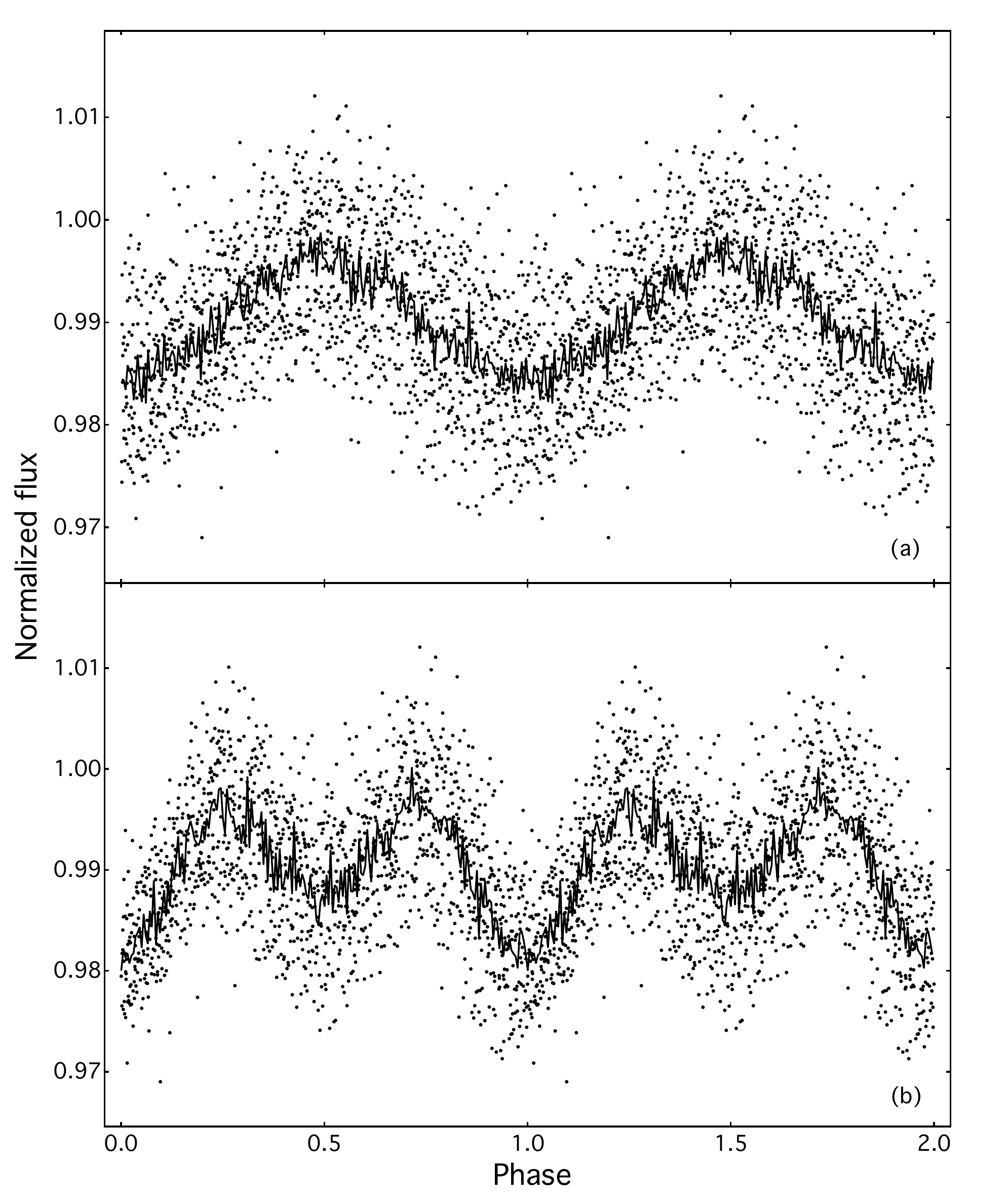}
\caption{Phased and binned light curves of \object{KZ\,Gem} are plotted by the scatter dots and solid lines, respectively. The K2 light curves folded on the period of 0.1112\,day (2.67\,hr) and of 0.2225\,day (5.34\,hr) were shown in panels (a) and (b), respectively.} \label{Figure 3}
\end{figure}

\begin{figure}
\centering
\includegraphics[width=9.0cm]{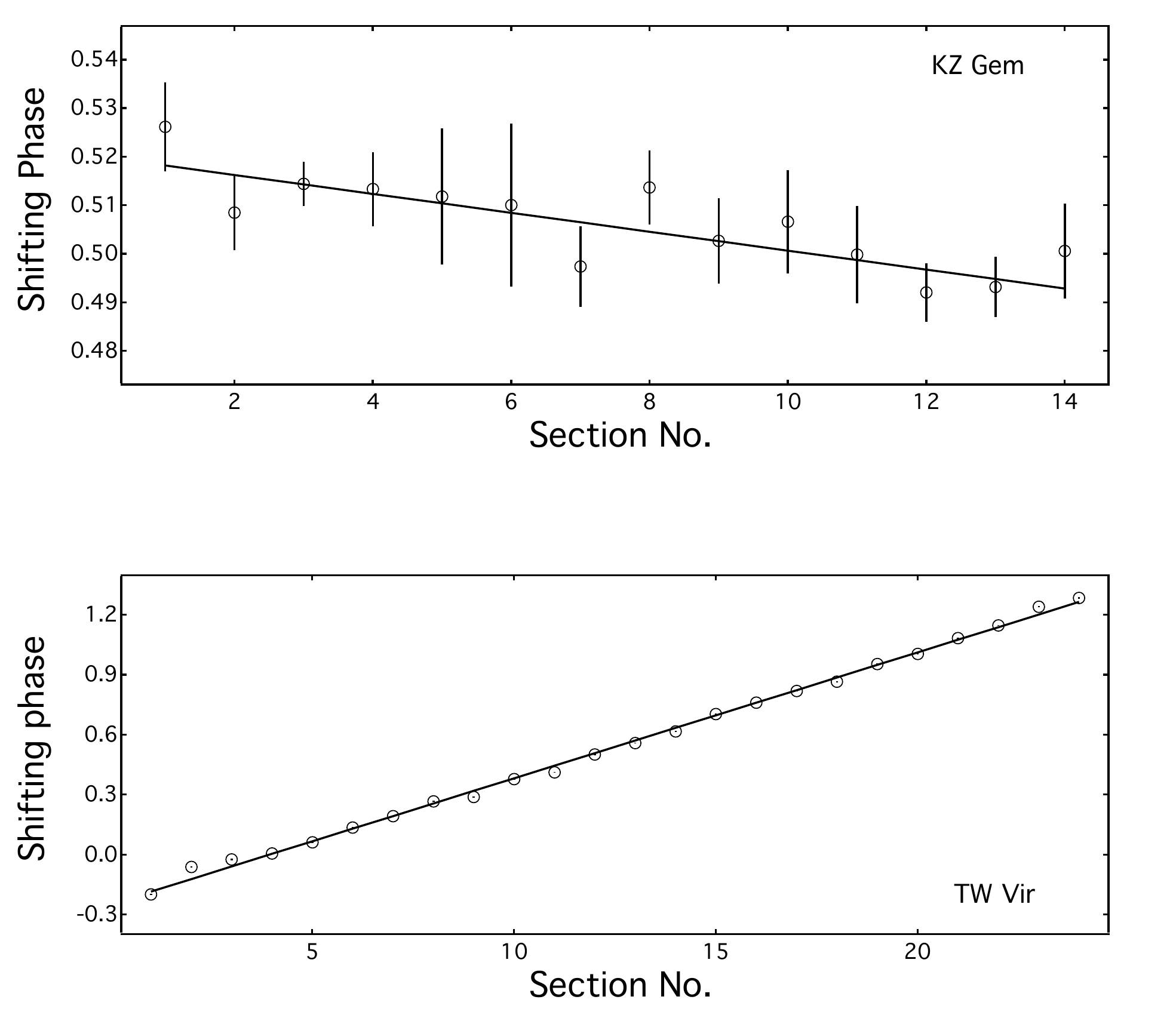}
\caption{The shifting phases calculated in each section of \object{KZ\,Gem} and \object{TW\,Vir} are plotted in the upper and lower panels, respectively. The solid lines denote the linear fit to the data.} \label{Figure 4}
\end{figure}

\begin{figure}
\centering
\includegraphics[width=9.0cm]{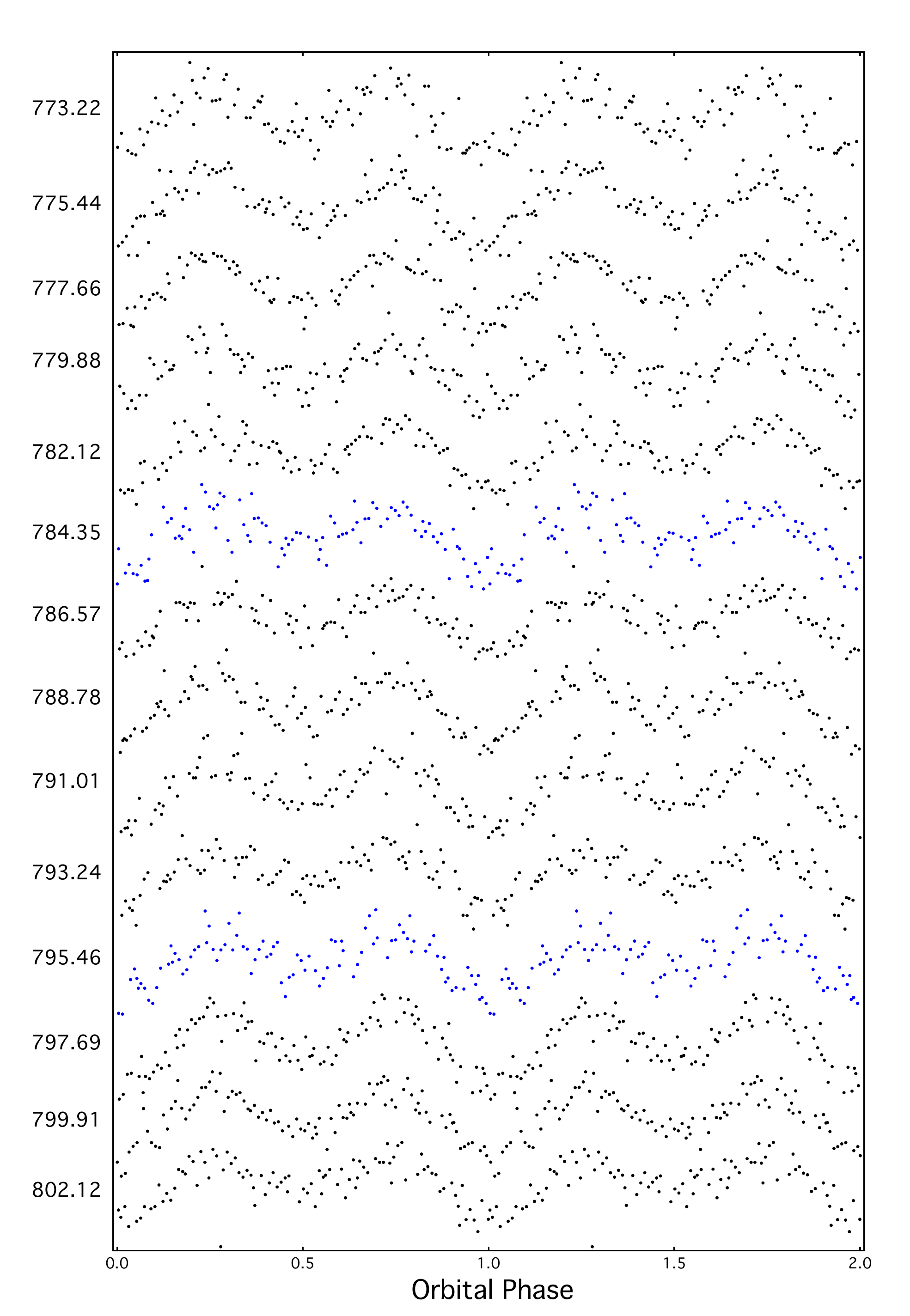}
\caption{Based on the 14 sections separated from the K2-C0 data of \object{KZ\,Gem}, all 14 phased light curves folded on the same period of 0.22242(1)\,day are stacked in chronological order from top to bottom. The labels of Y axis refer to the median times of the sections. The two blue sections indicate the unstable modulations.} \label{Figure 5}
\end{figure}

\begin{figure}
\centering
\includegraphics[width=9.0cm]{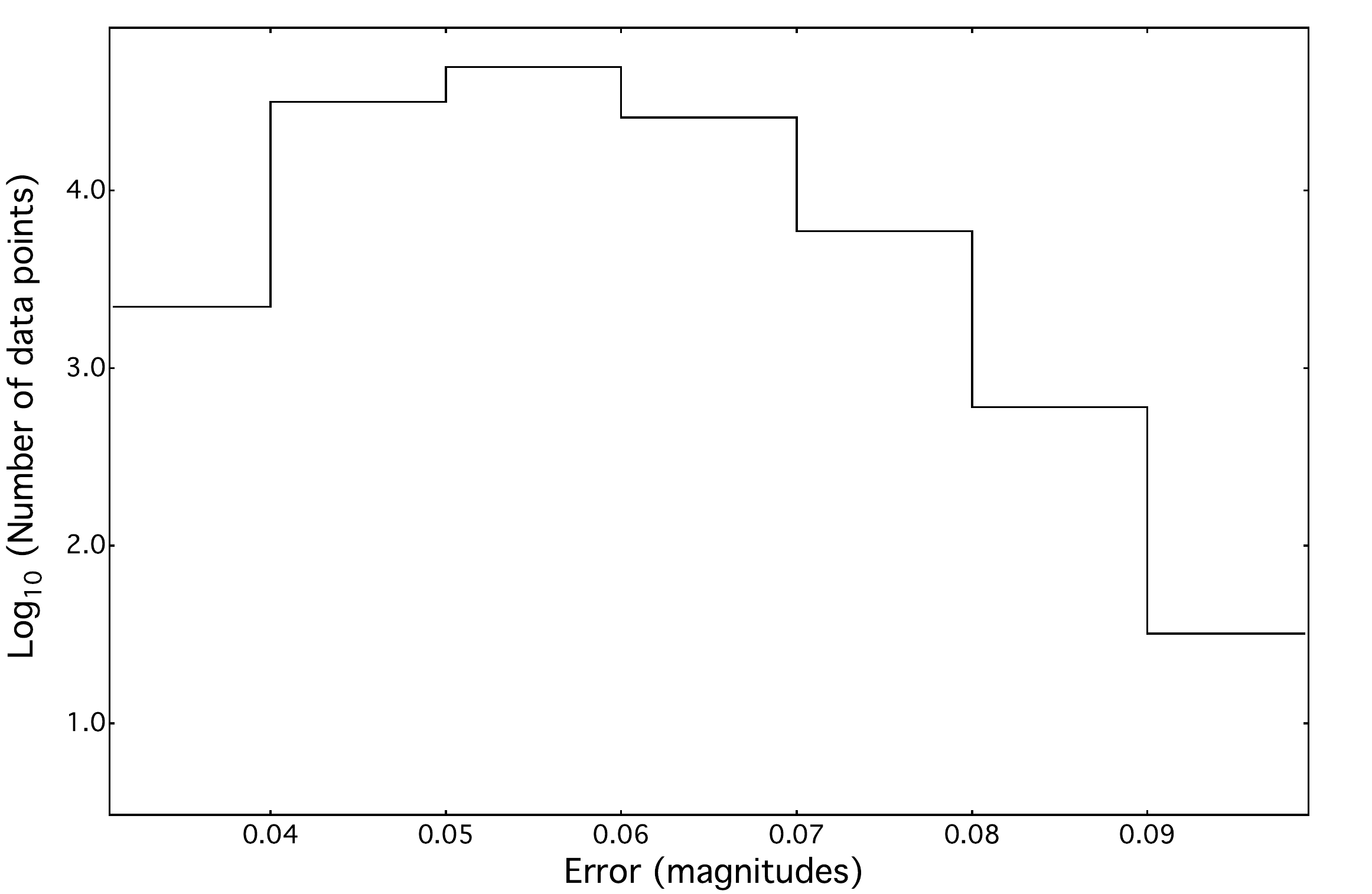}
\caption{The histogram of error distribution of the K2-C1 SC data for \object{RZ\,Leo}.} \label{Figure 6}
\end{figure}

\begin{figure*}
\centering
\includegraphics[width=14.0cm]{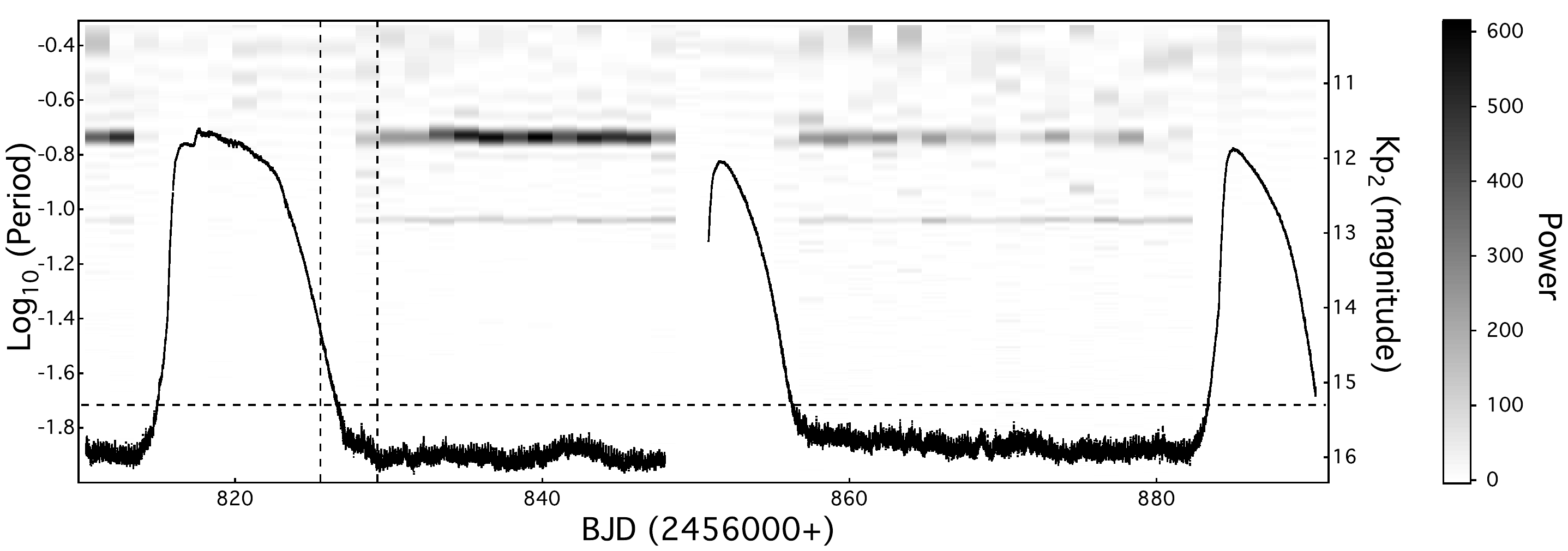}
\caption{Two-dimensional power spectrum of \object{TW\,Vir} calculated from a moving window 1.6\,day is overplotted with its K2 light curve in magnitudes. A horizontal dash line at Kp$_{2}$\,=\,15.3\,mag is arbitrarily used to distinguish the two luminosity states: outburst and quiescence. The light curve between the two vertical dash lines lasting $\sim$3.7\,day may denote the rebuilding process of the orbital modulation that was interrupted by the superoutburst.}\label{Figure 7}
\end{figure*}

\begin{figure}
\centering
\includegraphics[width=9.0cm]{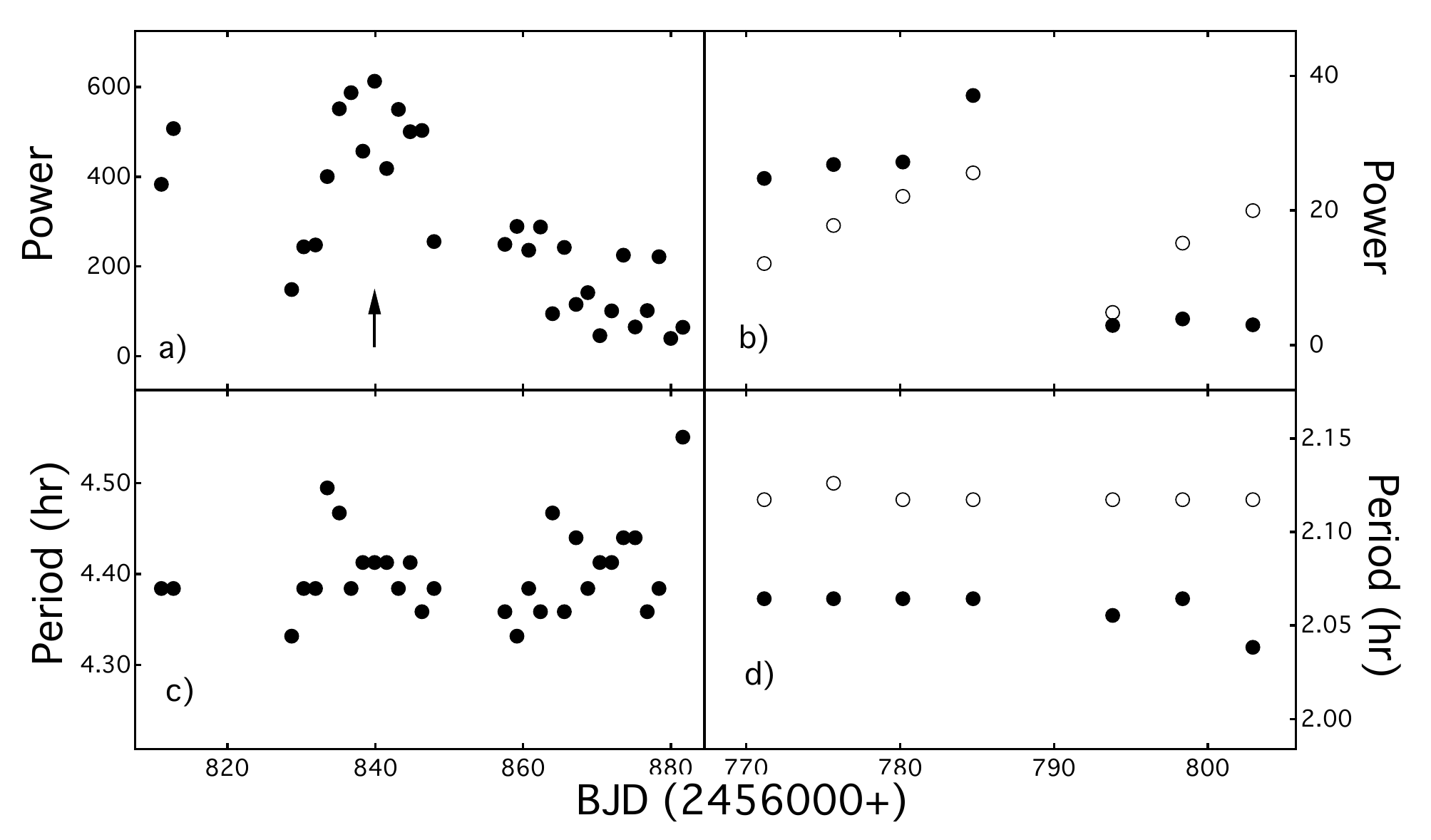}
\caption{The maximal power extracted from the 2-D power spectrum of \object{TW\,Vir} shown in Figure 7 and the corresponding periods were plotted in the left-side panels a) and c), respectively. Similarly, the two diagrams on the right-side are for \object{UV\,Gem}. The solid and open circles in the plots of \object{UV\,Gem} refer to the orbital and possibly remaining superhump periods after an unrecorded superoutburst \citep{dai17}, respectively. The arrow shown in panel a) denotes the changing point of the orbital modulation from strong to weak around BJD\,2456839.9.} \label{Figure 8}
\end{figure}

\begin{figure}
\centering
\includegraphics[width=9.0cm]{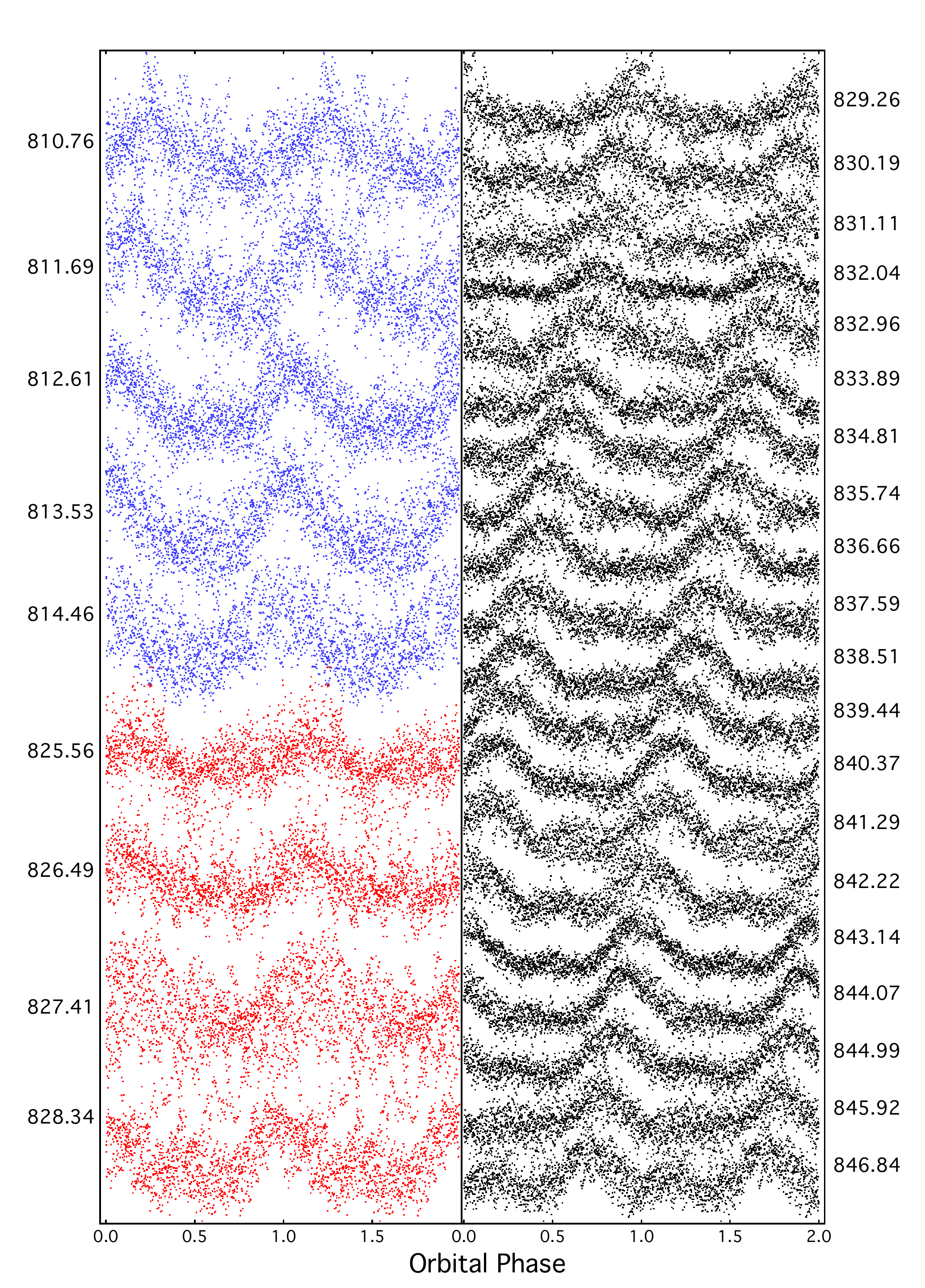}
\caption{The 29 phased light curves of \object{TW\,Vir} in quiescence around superoutburst folded on the initial period of 0.185\,day and stacked in time order. Like Figure 5, the median time of each section is labeled on the Y-axis. The blue and black light curves indicate the light curves before and after superoutburst, respectively. The red light curves are derived from the  part of light curve between the two vertical dash lines shown in Figure 7. According to our phase-correcting method, the 20 black light curves with the distinct double-hump modulation and the 4 red light curves were used to improve the orbital period of \object{TW\,Vir}.} \label{Figure 9}
\end{figure}

\begin{figure}
\centering
\includegraphics[width=9.0cm]{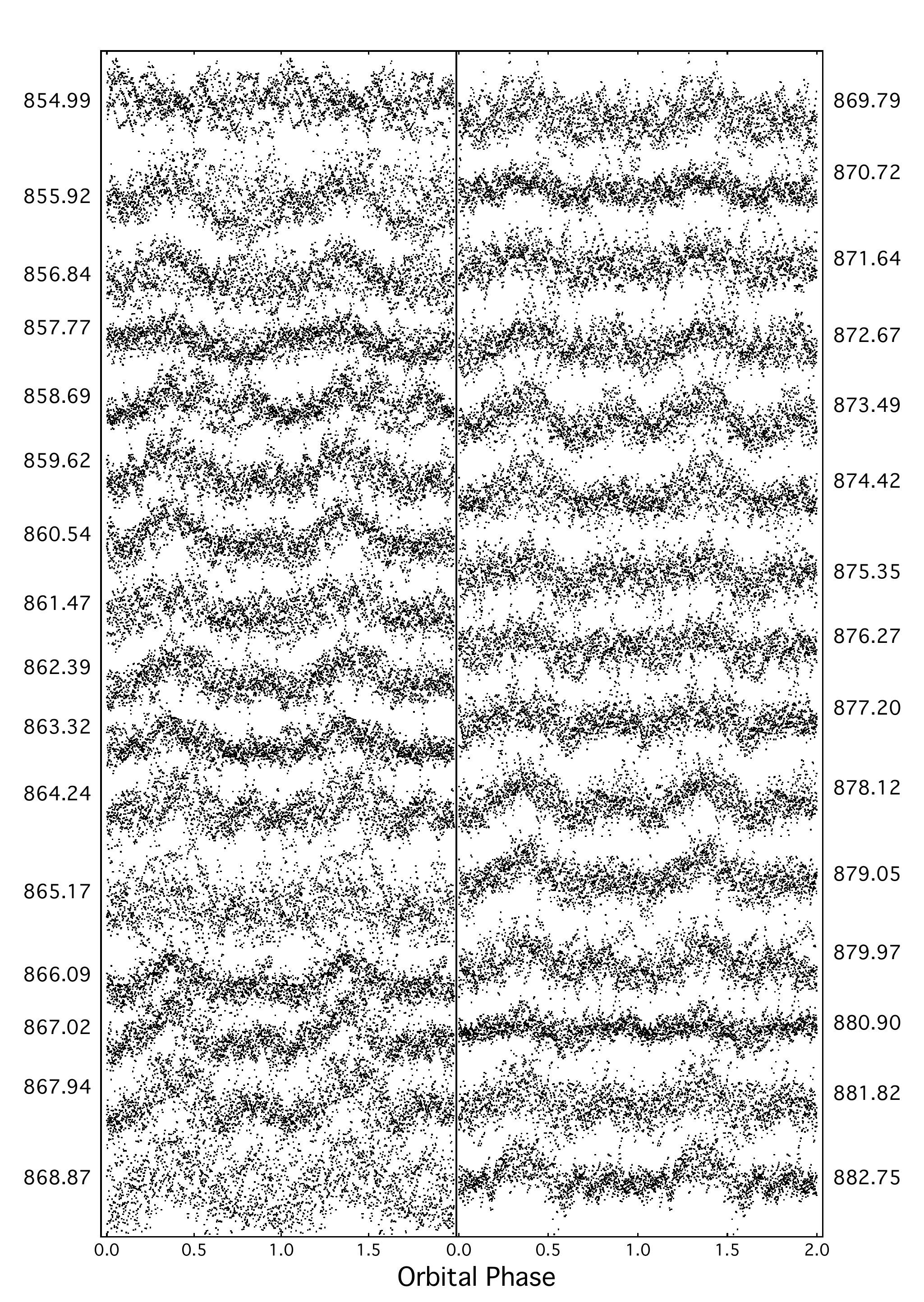}
\caption{The 31 phased light curves of \object{TW\,Vir} folded on the initial period of 0.185\,day between the two normal outburst are stacked in time order.} \label{Figure 10}
\end{figure}

\begin{figure}
\centering
\includegraphics[width=9.0cm]{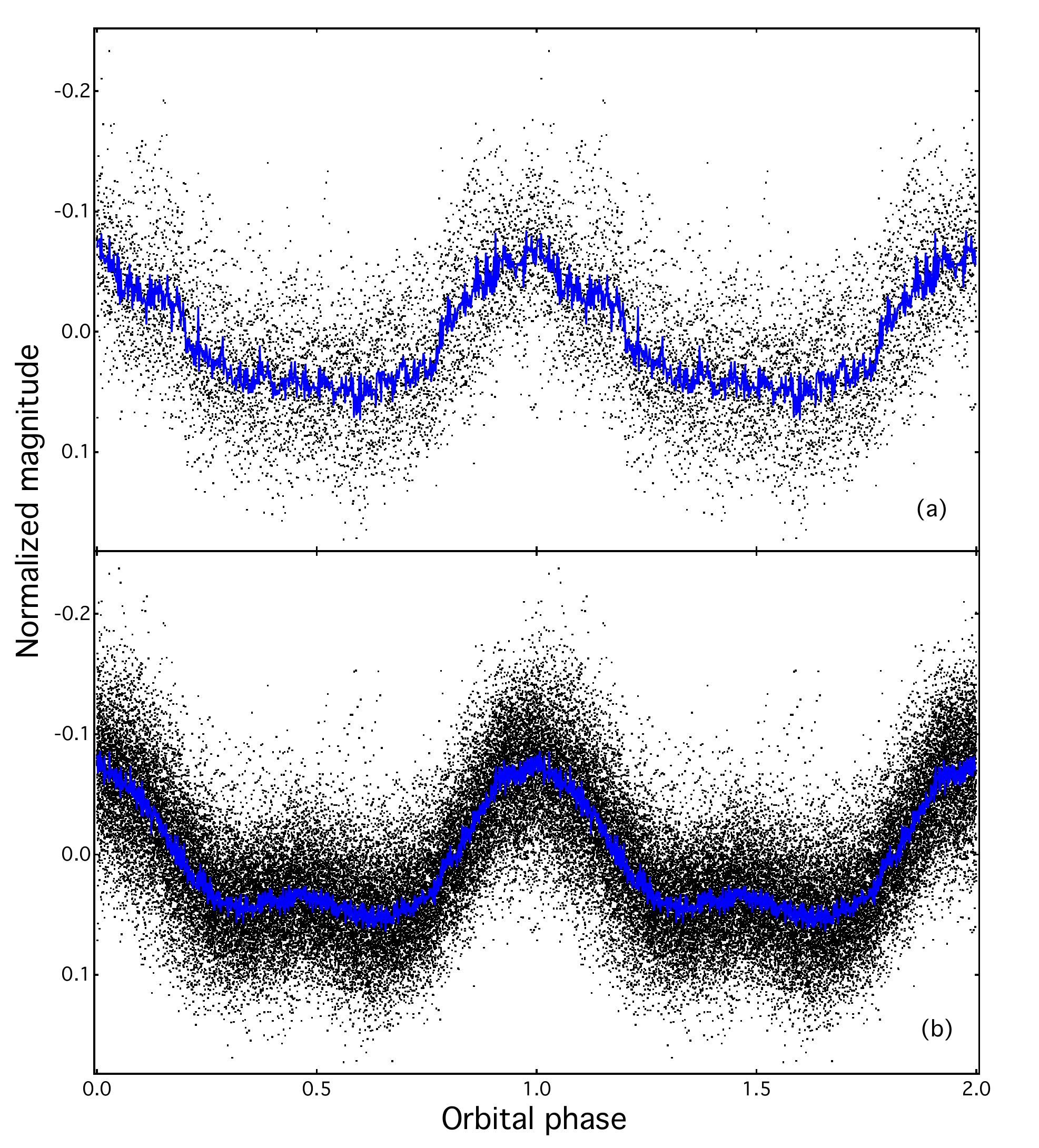}
\caption{Based on the light curves before and after superoutburst, the phased light curves of \object{TW\,Vir} with the corrected orbital period of 0.182682(3)\,day show the single-hump and double-hump modulation shown in panels (a) and (b), respectively. The blue solid lines are the binned light curves.} \label{Figure 11}
\end{figure}

\begin{figure}
\centering
\includegraphics[width=9.0cm]{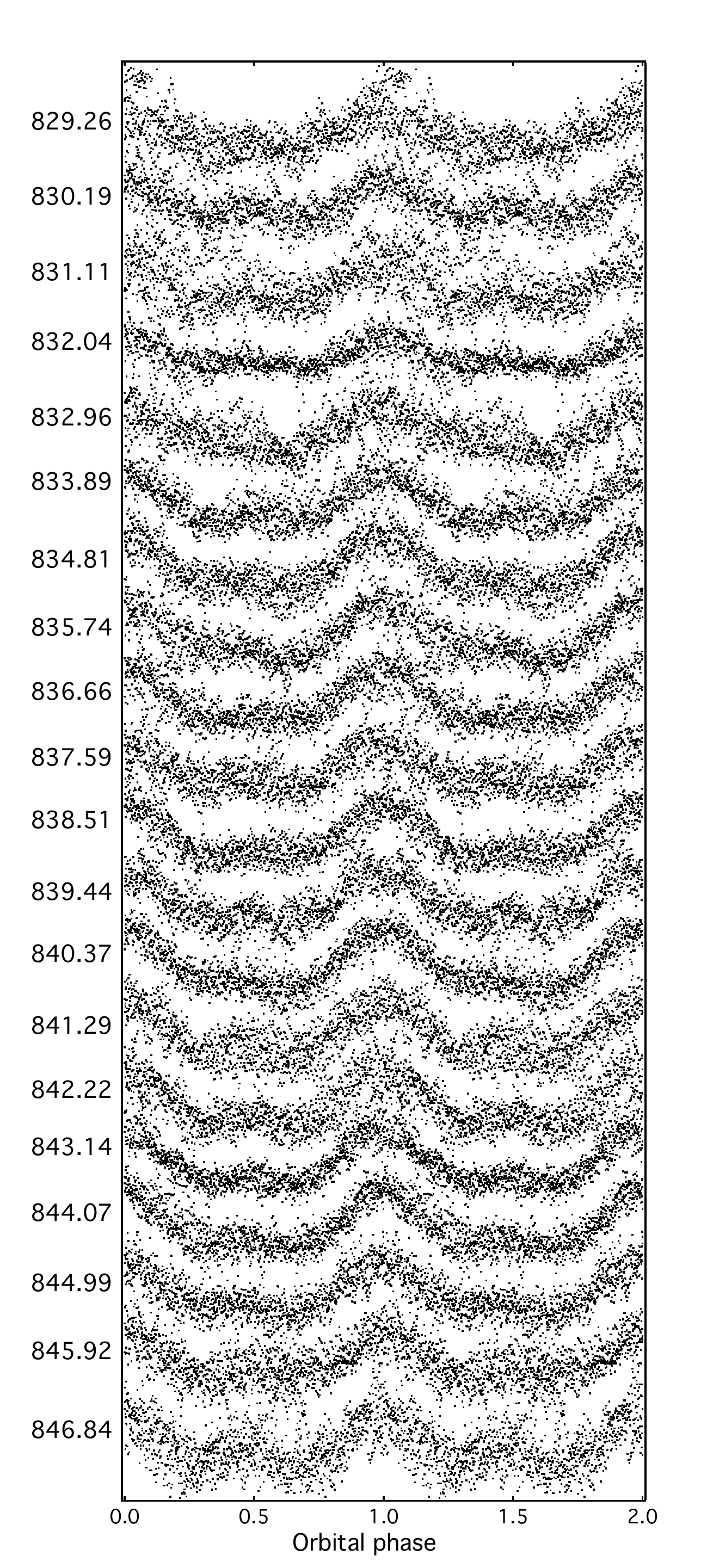}
\caption{The 20 new phased light curves of \object{TW\,Vir} with the corrected period of 0.182682(3)\,day are stacked in time order. All light curves are the same as those listed in the right-side panel of Figure 9. The significant and stable primary hump is set to be zero phase.} \label{Figure 12}
\end{figure}

\section{Results}

\subsection{\object{KZ\,Gem}}

\subsubsection{Period analysis}

\object{KZ\,Gem} is a poorly understood variable star first discovered by \cite{hof66} and \cite{kuk68}. Although it has been classified as a U Gem type dwarf nova (DN) in all versions of the CV catalogues \citep[e.g.][]{dow93,dow97,dow01,rk724}, there was not any formal report or literature supporting its DN subtype identification until a recent DN outburst occurred in 2015 January \citep{lan16}, which confirms its DN classification. The K2 variable catalogue$^{1}$ classified \object{KZ\,Gem} as 'OTHPER', i.e., other periodic and quasi-periodic variables, via a quick and automated period search \citep{arm15,arm16}. \cite{rk724} reported a period of 0.11122\,day as the orbital period. This orbital period has not been confirmed with spectra.

\footnotetext[1]{\scriptsize{http://archive.stsci.edu/k2/hlsp/k2varcat/search.php}}

In order to further test the period listed in the catalogs, the SFF corrected data of \object{KZ\,Gem} (Figure 1) was analyzed by using the LSP, the same method used in \cite{arm15}, and the PDM method. The strongest peak at a period of 0.1112\,day (2.67\,hr) is consistent with the orbital period listed in the updated CV catalogue (RKcat Edition 7.24 seen in \cite{rk724}). Note that there is also a peak with less power at the period of 0.2225\,day (5.34\,hr), which is exactly twice of the period of 2.67\,hr. The phased light curves of \object{KZ\,Gem} shown in the panels (a) and (b) of Figure 3 correspond to the periods of 2.67\,hr and 5.34\,hr, respectively. The phased light curve using a period of 2.67\,hr shows a single minimum per orbit. However, despite the large scatter of the raw data, the binned data using a period of 5.34\,hr shows two distinct minima in a single orbit. This means that the phased light curve with the period of 2.67\,hr actually overlaps the two different minima, and the data scatter blurs them into the apparent single minimum, which can be proved by the larger scatter in the panel (a) than that in the panel (b) of Figure 3. Thus, like  \object{1RXS\,J0632+2536} (J0632+2536) and \object{RZ\,Leo} \citep{dai16}, the orbital period of \object{KZ\,Gem} is likely 5.34\,hr and the 2.67\,hr peak is a harmonic. This newly determined period indicates that \object{KZ\,Gem} is a normal long-period DN rather than one of the CVs in the period gap. This light curve resembles the light curve caused by ellipsoidal modulation \citep[e.g][]{boc79}.

A higher precision orbital period can be obtained by using our phase-correcting method. The K2 time series of \object{KZ\,Gem} spanning over 30 days was separated into 14 sections, using the period of 5.34\,hr and setting N$_{s}$ as 10 (as the data is in LC model). Using Equation (1), the final orbital period of \object{KZ\,Gem} is calculated to be 0.22242(1)\,day based on the shift rate S$_{phase}$=\,-0.002(5) derived from the 14 shifting phases shown in the upper panel of Figure 4. The precision of the orbital period is an order of magnitude higher than that of the starting period obtained from the LSP and the PDM methods. In order to find further evidence that this is the orbital period of \object{KZ\,Gem}, the SFF corrected data of \object{KZ\,Gem} were first converted into K2 magnitude Kp$_{2}$ using the conversion in \cite{dai16}, and then all the 14 phased light curves folded on the improved orbital period of 0.22242(1)\,day, with an average amplitude of 0.0025\,mag, were stacked (Figure 5). These light curves indicate that the orbital modulation of \object{KZ\,Gem} is basically stable except for the two sections around BJD\,2456784.35 and BJD\,2456795.46. Obviously, all of the stable light curves with the smaller scatter show detailed features of the modulation including two different minima and similar maxima in one orbit, which is the typical ellipsoidal modulation light curve found in other binary systems, such as \object{Nova A0620-00} \citep{mcc83} and \object{TT\,Crt} \citep{szk92}.

\subsubsection{Light-curve morphology}

The stability of the orbital modulation of \object{KZ\,Gem} is also evident from the flux ratios (Figure 2). All four diagrams in Figure 2 are not only very close to the dashed line of 1, but also show considerably flat patterns with slopes around 10$^{-5}$, which means that this low-amplitude modulation is stable. In spite of the scatter, the linear fit shown in the plot of R$_{peak}$ is closer to the line of 1 than that in the plot of R$_{dip}$. This may provide support that the orbital period of \object{KZ\,Gem} is 0.22242(1)\,day and the corresponding orbital modulation has two minima per orbital cycle rather than a single sinusoidal-like oscillation (as shown in the (b) and (a) panels of Figure 3).

In order to determine whether the stability of the orbital modulations of \object{KZ\,Gem} is unusual, we also investigated the stability of three other low-inclination CV type light curves: \object{J0632+2536} observed in K2-C0, and \object{RZ\,Leo} and the pre-CV \object{WD\,1144+011} observed in K2-C1. Note that these CVs never show any superoutburst during the K2 observations. By using the four flux ratios defined in Section 3.2, their flat patterns like \object{KZ\,Gem} shown in Figure 2 illustrate that all three CV light curves with the distinct orbital modulations derived by \cite{dai16} are basically stable throughout the K2 observations. Although \object{J0632+2536} shows negative slopes in R$_{peak}$ and A$_{pri}$ and a quadratic-like pattern in R$_{dip}$, there is only a small distortion of the double-hump orbital modulation as the K2-C0 observation progresses. In addition, the sharp decrease of system light of \object{J0632+2536} occurring around BJD\,2456790 shown in the top panel of Figure 1 of \cite{dai16} is coincident with the change in R$_{dip}$, as pointed out by the arrow in the R$_{dip}$ plot. For \object{RZ\,Leo}, the fairly constant orbital modulation is in contrast to a model of a shifting hot spot proposed by \cite{men01}. And the "anti-humps" claimed by \cite{men99} were undetected in the K2-C1 SC data. Since the average error of \object{RZ\,Leo} is from 0.04\,mag to 0.07\,mag and the maximal error bar is $<$0.1\,mag (Figure 6), the uncertainties of the data points are not enough to explain the large amplitude dispersion $\sim$\,0.43\,mag shown in the phased light curve of \object{RZ\,Leo}. Therefore, the flat patterns of \object{RZ\,Leo} imply that this scatter amplitude results from a uniform drift of the system light on a short timescale, rather than from any variations in the disk or hot spot. 

\subsection{\object{TW\,Vir}}

\subsubsection{Period analysis}

Although \cite{sha83} has published a spectroscopic orbital period of 4.38\,hr for \object{TW\,Vir} and there were 7872 photometric observations from 1955 to 1997 \citep{ak02}, the existence of a photometric orbital modulation is still unknown. The recent period analysis during the plateau of a superoutburst in \object{TW\,Vir} \citep{dai16} only showed the well-known 6\,hr period due to the K2 satellite pointing drift, and some ambiguous peaks smaller than 4.38\,hr, which could be harmonics of this pseudo 6-hr period. The online K2 variable catalogue$^{1}$ classified \object{TW\,Vir} as 'RRab' (i.e., RRab lyrae star) without any available phase-folded light curve.

\footnotetext[1]{\scriptsize{http://archive.stsci.edu/k2/hlsp/k2varcat/search.php}}

A 2-D power spectrum analysis with a moving window of 1.6\,day was applied to the whole K2 SC light curve of \object{TW\,Vir}. Inspection of Figure 7 illustrates two significant and coherent black bars around the periods of 0.183(4)\,day (4.4\,hr) and 0.092(4)\,day (2.2\,hr), respectively. The latter period with less power is the second harmonic of the former one. This 2-D power spectrum diagram supports the previous orbital period derived by \cite{sha83}, and shows that the orbital modulation of \object{TW\,Vir} found in quiescence disappears during outburst (a result similar to that found for \object{UV\,Gem} \citep{dai17}). Thus, we focused on the quiescent light curve to improve the precision of the orbital period. Based on the times of appearance and disappearance of the orbital modulation, the light curve when Kp$_{2}$ was fainter than 15.3\,mag was defined to be the quiescent light curve. This light curve covers a total of 25 days, separated into three quiescent sections by one superoutburst and two normal outbursts (Figure 7). The latter two quiescent sections completely cover the intervals between the three outbursts. The orbital modulation in quiescence between the superoutburst and the following normal outburst is remarkably stronger than that between the two normal outbursts. The maximal powers and the corresponding periods tracking along the black bar of the 2-D power spectrum are plotted in the upper and lower panels of Figure 8, respectively.

Considering the width of the black bar shown in Figure 7 and the dispersion of the orbital period around 0.008\,day ($\sim$\,0.2\,hr) denoted by the irregular variation shown in panel c) of Figure 8, we needed a detailed period analysis to improve the precision of the orbital period of \object{TW\,Vir}. We accomplished this by using the previously described phase-correcting method. An initial period was intentionally set to be 0.185\,day (4.44\,hr) with a small deviation $\sim$4\,min away from the average period 4.38\,hr shown in the panel c) of Figure 8. Based on this initial period and the preset N$_{s}$\,=\,5, the SC light curve of \object{TW\,Vir} observed in K2-C1 was separated into 80 sections. After subtracting the sections during the outbursts, the resulting 60 sections in quiescence were stacked in Figure 9 and 10 illustrating the quiescent light curves around the superoutbust and between the two normal outbursts, respectively. The quiescent orbital modulations of \object{TW\,Vir} around the superoutburst are more stable despite the apparent differences in the modulation profiles before and after superoutburst (Figure 11). We found that a stable orbital modulation was not established until BJD\,2456829.26. There are 24 sections from BJD\,2456825.56 to BJD\,2456846.84 that were analyzed by our phase-correcting method. The orbital period of \object{TW\,Vir} was improved to 0.182682(3)\,day based on the linear shift phases shown in the lower panel of Figure 4. The precision of the derived orbital period is an order magnitude higher than that obtained by \cite{sha83} and three orders of magnitude higher than that derived from the 2-D power spectrum analysis. The 20 sections from BJD\,2456829.26 to BJD\,2456846.84 were folded on this period, and the resulting phased light curves are shown in Figure 12. These phased light curves show clear orbital modulations. However, an accurate ephemeris for \object{TW\,Vir} cannot be derived from the K2 data since the cycle-to-cycle scatter makes a precise epoch difficult to determine.

\subsubsection{Light-curve morphology}

The phased light curves before and after superoutburst show different shapes (Figure 12). During most quiescence times, a double hump orbital modulation (panel b) is apparent. However, the orbital modulation before the superoutburst appears to gradually mimic the shape of a typical superhump profile with a single maximum and minimum per cycle, similar to many other SU\,UMa type DNe \citep[e.g.][]{pat03,kat13a,kat13b}. Panel (a) of Figure 11 illustrates this superhump-like modulation profile. Since \object{TW\,Vir} is classified as a UG Gem-type DN \citep{oco32} and its orbital period is far longer than the typical period of the SU\,UMa type DN, this single-peak modulation may represent the accretion pattern caused by an increased hot spot area before the upcoming superoutburst. The short time base of the K2 data before superoutburst cannot totally rule out the other possibility that the second hump may be overwhelmed by the large scatter. Post superoutburst, an orbital modulation began to form after BJD\,2456825.56, which is close to the half time of the decay section of the outburst light curve. This part of the light curve from BJD\,2456825.56 to BJD\,2456846.84, lasting $\sim$3.7\,day at the end of the decay from superoutburst (Figure 7), was divided into 4 sections shown in the bottom left plots of Figure 9. These sections clearly indicate that the orbital modulation is becoming stronger, implying a reconstruction stage of the orbital modulation that was disrupted by the superoutburst. Panel a) of Figure 8 shows that the amplitude of the orbital modulation after the superoutburst first experiences a fast strengthening stage lasting about 10 days and then continuously weakens. This behavior is compared to that of \object{UV\,Gem} in Figure 8b.

Since the quiescence following the next normal outburst seems to give rise to a chaotic orbital modulation, it is pointless to phase the entire K2-C1 quiescent dataset using this orbital period. Instead, we restrict our analysis to data taken between the superoutburst and following normal outburst(BJD\,2456829-2456847). The relative high stability and the long time base of the light curve between the superoutburst and the following normal outburst (i.e., from BJD\,2456829 to BJD\,2456847), allowed this section to be used for further analysis. Since the indistinct second humps shown in panel (b) of Figure 11 are highly variable, A$_{pri}$ is the only reliable parameter that could be used. The plot of A$_{pri}$ in Figure 2 for \object{TW\,Vir} indicates a quadratic-like variation in primary amplitude.

The turning point of A$_{pri}$ is basically in accord with the strongest orbital modulation shown in panel a) of Figure 8. Combined with the variations in the amplitude of the spectrum power, the quadratic-like variation in A$_{pri}$ implies a scenario where the orbital modulation of \object{TW\,Vir} is quickly strengthened after superoutburst, lasts about 10 days, and then turns to a decay stage before the next normal outburst. Based on the quadratic fit to the curve shown in the A$_{pri}$ plot of Figure 2, the spread of primary amplitude during this part of quiescence from superoutburst to normal outburst is about 4\%. Since the enhancement of the primary amplitude is accompanied by the appearance of the orbital modulation before BJD\,2456839.9, the enlarged amplitude may imply an adjustment of the accretion disk structure after superoutburst. Nevertheless, this recovered orbital modulation generally becomes unstable after BJD\,2456839.9, and then almost falls into chaos after the following normal outburst, which can be clearly seen in Figure 10. The transformation of the orbital modulation in a low inclination CV from stable to unstable has not been reported in the past. Since the turning point appears in the middle of the outburst interval, it is apparent that a stable orbital modulation, based on a typical CV accretion model \citep{war03}, is not maintained. Changes in accretion can definitely give rise to variations in orbital modulation, e.g., as recently reported in the eclipsing SU\,UMa type DN \object{V1239\,Her} \citep{gol15}. Thus, the observed decrease of the amplitude of the orbital modulation suggests changes in the accretion disk structure of \object{TW\,Vir} around BJD\,2456839.9. The high instability of the orbital modulation after the normal outburst may naturally explain why the optical orbital modulation of \object{TW\,Vir} was not observed in the 7872 observations of \cite{ak02}.

\section{Conclusions}

The long datasets of K2-C0 and K2-C1 have allowed refined orbital periods for \object{KZ\,Gem} and \object{TW\,Vir} as well as a study of the stability of their photometric orbital variations. A phase-correcting method was successfully used to improve the orbital period of \object{KZ\,Gem} to 0.22242(1)\,day. This period confirms that \object{KZ\,Gem} is not a CV in the period gap, but a long-period DN. Our phased light curve folded on this period shows typical ellipsoidal modulations. Radial velocity measurements can provide further confirmation. By analyzing four defined flux ratios: R$_{peak}$, R$_{dip}$, A$_{pri}$  and A$_{sec}$ and creating colormaps, \object{KZ\,Gem}, \object{J0632+2536}, \object{RZ\,Leo} and \object{WD\,1144+011} are shown to have flat or sloping patterns that indicate stable orbital modulations during quiescence. Despite the large scatter and variability of CV light curves, the orbital modulation is commonly stable for these low inclination CVs. However, \object{TW\,Vir} with a superoutburst and two normal outbursts shows unstable orbital modulation hidden in quiescence. Removing the outburst times from the K2-C1 SC data reveals its variable orbital modulation. By applying a phase-correcting method to 24 sections of data with the highest stability spanning over 17 days, the orbital period of \object{TW\,Vir} is improved to 0.182682(3)\,day. Two types of orbital modulation with signal-hump and double-hump profiles are derived before and after superoutburst, respectively. The amplitude of the orbital modulations are much stronger after a superoutburst than after the following normal outburst. These data may be the first visible detection of the rebuilding process of the accretion disk structure following a superoutburst. The reconstruction period of the orbital modulation is estimated as $\sim$3.7\,day. Based on a part of the quiescent light curve of \object{TW\,Vir} with a stable double-hump modulation, a nonlinear variation exists in the plot of A$_{pri}$ for \object{TW\,Vir}, corresponding to the time when the orbital modulation amplitude changes from strong to weak, rather than to any detectable transient event. The changing amplitudes are indicative of a changing accretion disk structure from superoutburst to quiescence. 

\begin{acknowledgements}

This work was partly supported by CAS Light of West China Program and the Science Foundation of Yunnan Province (No. 2016FB007). PS acknowledges support from NSF grant AST-1514737. MRK and PG acknowledge support from the Naughton Foundation and the UCC Strategic Research Fund.

\end{acknowledgements}


\end{CJK*}

\begin{thebibliography}{}

\bibitem[Ak et al. (2002)]{ak02}
  Ak, T., Ozkan, M. T., \& Mattei, J. A. 2002, A\&A, 389, 478.
\bibitem[Armstrong et al. (2015)]{arm15}
  Armstrong, D. J., Kirk, J., Lam, K. W. F., McCormac, J., Walker, S. R., et al. 2015, A\&A, 579, 19.
\bibitem[Armstrong et al. (2016)]{arm16}
  Armstrong, D. J., Kirk, J., Lam, K. W. F., McCormac, J., Osborn, H. P., et al. 2016, MNRAS, 456, 2260.
\bibitem[Bochkarev et al. (1979)]{boc79}
  Bochkarev, N. G., Karitskaya, E. A., \& Shakura, N. I. 1979, Pisma v Astronomicheskii Zhurnal, 23, 8.
\bibitem[Bruch (1991)]{bru91}
  Bruch, A. 1991, A\&A, 251, 59.
\bibitem[Bruch (1992)]{bru92}
  Bruch, A. 1992, A\&A 266, 237.
\bibitem[Dai et al. (2016)]{dai16}
  Dai, Z. B., Szkody, P., Garnavich, P. M., \& Kennedy, M. R. 2016, AJ, 152, 5.
\bibitem[Dai et al.(2017)]{dai17}
  Dai, Z. B., Szkody, P., Garnavich, P. M., \& Kennedy, M. R. 2017, PoS-SISSA, 28, in press.
\bibitem[Downes \& Shara (1993)]{dow93}
  Downes, R. A., \& Shara, M. M. 1993, PASP 105, 127. 
\bibitem[Downes et al. (1997)]{dow97}
  Downes, R. A., Webbink, R. F., \& Shara, M. M. 1997, PASP, 109, 345. 
\bibitem[Downes et al. (2001)]{dow01}
  Downes, R. A., Shara, M. M., Ritter, H., Kolb, U., et al. 2001, PASP, 113, 764.
\bibitem[Golysheva et al. (2015)]{gol15}
  Golysheva, P., Shugarov, S., Katysheva, N., \& Khruzina, T. 2015, ASPC, 496, 231.
\bibitem[Howell et al. (2014)]{how14}
  Howell, S. B., Sobeck, C., Haas, M., et al. 2014, PASP, 126, 398.
\bibitem[Hoffmeister (1966)]{hof66}
  Hoffmeister, C. 1966, AN, 289, 139.
\bibitem[Kato \& Osaki (2013a)]{kat13a}
  Kato, T., \& Osaki, Y. 2013a, PASJ, 65, 97.
\bibitem[Kato et al. (2013b)]{kat13b}
  Kato, T., Hambsch, Franz-Josef, Maehara, H., Masi, G., Miller, I. 2013b, PASJ, 65, 23.
\bibitem[Kennedy et al. (2016)]{ken16}
  Kennedy, M. R., Callanan, P., Garnavich, P. M., Szkody, P., Bouanane, S., et al. 2016, MNRAS, 459, 3622.
\bibitem[Kukarkin et al. (1968)]{kuk68}
  Kukarkin, B. V., Kholopov, P. N., Efremov Y. N., Kurochkin N. E., Frolov M. S., et al. 1968, IBVS, 311, 1.
\bibitem[Lange (2016)]{lan16}
  Lange, T. 2016, BAVSR, 65, 45.
 \bibitem[Lomb (1976)]{lom76}
  Lomb, N. 1976, Ap\&SS, 39, 447.
\bibitem[McClintock et al. (1983)]{mcc83}
  McClintock, J. E., Petro, L. D., Remillard, R. A., \& Ricker, G. R. 1983, ApJ, 266, L27.
\bibitem[Mennickent et al. (1999)]{men99}
  Mennickent, R. E., Sterken, C., Gieren, W. \& Unda, E. 1999, A\&A, 352, 239.
\bibitem[Mennickent \& Tappert (2001)]{men01}
  Mennickent, R. E., \& Tappert, C. 2001, A\&A, 372, 563.
\bibitem[O$'$Connell(1932)]{oco32}
  O$'$Connell, D, J. K. 1932, Harvard College Observatory Bulletin, 890, 18.
\bibitem[Patterson et al. (2003)]{pat03}
  Patterson, J., Thorstensen, J. R., \& Kemp, J., et al. 2003, PASP, 115, 1308.
\bibitem[Ritter \& Kolb (2003)]{rk724}
  Ritter, H., \& Kolb, U. 2003, A\&A, 404, 301.
\bibitem[Scargle (1982)]{sca82}
  Scargle, 1982, ApJ, 263, 835.
\bibitem[Shafter (1983)]{sha83}
  Shafter, A. W. 1983, IBVS, 2377.
\bibitem[Stellingwerf (1978)]{ste78}  
  Stellingwerf, R. F. 1978, ApJ, 224, 953.
\bibitem[Still \& Barclay (2012)]{sti12}
  Still, M., \& Barclay, T. 2012, ascl soft, ascl:1208, 004.
\bibitem[Szkody et al. (1992)]{szk92}
  Szkody, P., Williams, R. E., Margon, B., Howell, S. B., \& Mateo, M. 1992, ApJ, 387, 357.
\bibitem[Szkody et al. (2016)]{szk16}
  Szkody, P., Mukadam, A. S., G\"{a}nsicke, B. T., Toloza, O., Dai, Z.-B., et al. 2016, AJ, submitted.
\bibitem[Taylor et al. (1999)]{tay99}
  Taylor, C. J., Thorstensen, J. R., \& Patterson, J. 1999, PASP, 111, 184.
\bibitem[Vanderburg \& Johnson (2014)]{van14a}
  Vanderburg, A. \& Johnson, J. A. 2014, PASP, 126, 948.
\bibitem[Vanderburg (2014)]{van14b}
  Vanderburg, A. 2014, arXiv, 1412, 1827.
\bibitem[Warner (1995)]{war95}  
 Warner, B. 1995, Cataclysmic Variable Stars. (Cambridge Univ. Press, Cambridge)
\bibitem[Warner (2003)]{war03}
  Warner, B. 2003, Cataclysmic Variable Stars. (Cambridge Univ. Press, Cambridge)
  
\end{thebibliography}
\end{document}